\documentclass[%
reprint,
 amsmath,amssymb,
 aps,
 prd,
 10pt,
]{revtex4-2}

\usepackage{graphicx}% Include figure files
\usepackage{dcolumn}% Align table columns on decimal point
\usepackage{bm}% bold math
\usepackage{hyperref}% add hypertext capabilities
\usepackage{subfig}
\usepackage{ragged2e}
\usepackage[compat=1.1.0]{tikz-feynman}
\usepackage{color}
\definecolor{comment}{rgb}{.4,.4,.4}
\usepackage{algorithm}
\usepackage{subcaption}
\usepackage[
noEnd=false,
indLines=false,
italicComments=false,
rightComments=false,
commentColor=comment,
beginComment={~~//~},
endComment={},
]{algpseudocodex}
\algrenewcommand\textproc{\textrm}

\begin{document}

\preprint{APS/123-QED}

\title{Constraints on SMEFT operators from $Z \to \mu \mu bb$ decay}%
%\thanks{A footnote to the article title}

\author{Zijian \surname{Wang}}%
\email{wangzijian@stu.pku.edu.cn}%
\author{Tianyi \surname{Yang}}%
%\email{wangzijian@stu.pku.edu.cn}%
\author{Tianyu \surname{Mu}}%
%\email{wangzijian@stu.pku.edu.cn}%
\author{Andrew \surname{Levin}}%
%\email{wangzijian@stu.pku.edu.cn}%
\author{Qiang \surname{Li}}%
%\email{qliphy0@pku.edu.cn}%
\affiliation{School of Physics and State Key Laboratory of Nuclear Physics and Technology, Peking University, Beijing, 100871, China}%

\date{\today}% It is always \today, today,
             % but any date may be explicitly specified

\begin{abstract}

The Standard Model Effective Field Theory (SMEFT) provides a systematic framework to parametrize indirect effects of heavy new physics in precision measurements. In this work, we study the \ensuremath{\mu^+\mu^-b\bar b} final state in the reconstructed \ensuremath{Z}-pole region and derive constraints on selected dimension-six SMEFT operators involving second-generation leptons and third-generation quarks. Signal and background processes are simulated at leading order using standard Monte Carlo tools, followed by parton showering and fast detector simulation with simplified \ensuremath{b}-tagging effects. We focus on four-fermion operators involving leptons and bottom quarks, as well as operators modifying effective \ensuremath{Z}--fermion couplings. The dependence of the selected event yield on individual Wilson coefficients is obtained using SMEFT reweighting and parametrized including both interference and quadratic dimension-six contributions. Expected constraints are derived with an Asimov likelihood-ratio approach for integrated luminosities of \ensuremath{138~\mathrm{fb}^{-1}} and \ensuremath{3000~\mathrm{fb}^{-1}}. The results provide channel-specific constraints on flavor-resolved SMEFT interactions in mixed leptonic--hadronic \ensuremath{Z}-pole final states and offer a useful reference for future analyses including higher-order and systematic effects.

\end{abstract}

\maketitle

%\tableofcontents

\section{Introduction}

The $Z$ boson has long served as a cornerstone of precision tests of the
electroweak sector of the Standard Model (SM). At both the LEP and the LHC, measurements of $Z$-boson production and decay have reached a precision that allows indirect probes of physics beyond the SM \cite{ALEPH:2005ab,ATLAS:2016nqi,CMS:2019zmd}. In particular, rare and multi-body $Z$-boson decays probe kinematic regimes and operator structures not fully constrained by inclusive observables.

The Standard Model Effective Field Theory (SMEFT) offers a model-independent framework to parameterize the effects of heavy new physics through higher-dimensional operators constructed from SM fields \cite{Buchmuller:1985jz,Grzadkowski:2010es,Brivio:2017vri}. In recent years, SMEFT interpretations of $Z$-boson observables have attracted considerable attention. Purely leptonic decay modes, such as $Z \to 4\ell$, provide stringent constraints on electroweak and four-lepton operators, while complementary information has been obtained from SMEFT analyses of inclusive $Z$ production and $Z$+jets, probing quark electroweak couplings and dipole interactions \cite{Boughezal:2020uwq,Falkowski:2015krw}.

Despite this progress, $Z$-boson decays into mixed leptonic--hadronic final
states remain relatively unexplored in the SMEFT context. Such channels are sensitive to a distinct set of operators, in particular four-fermion contact interactions involving both leptons and quarks, as well as flavor-dependent modifications of $Z$--fermion couplings \cite{Berthier:2015oma}. Among these, the $\mu^+\mu^-b\bar b$ final state in the $Z$-pole region is of special interest. The presence of bottom quarks enhances sensitivity to operators coupling to the third quark generation, while the dimuon final state allows for precise reconstruction of the leptonic kinematics. At the same time, the analysis poses experimental challenges due to sizable QCD backgrounds and the reliance on heavy-flavor tagging \cite{CMS:2019mwi,ATLAS:2018kcu}.

\begin{figure}[H]
	\centering
	\includegraphics
	[width=0.35\textwidth]{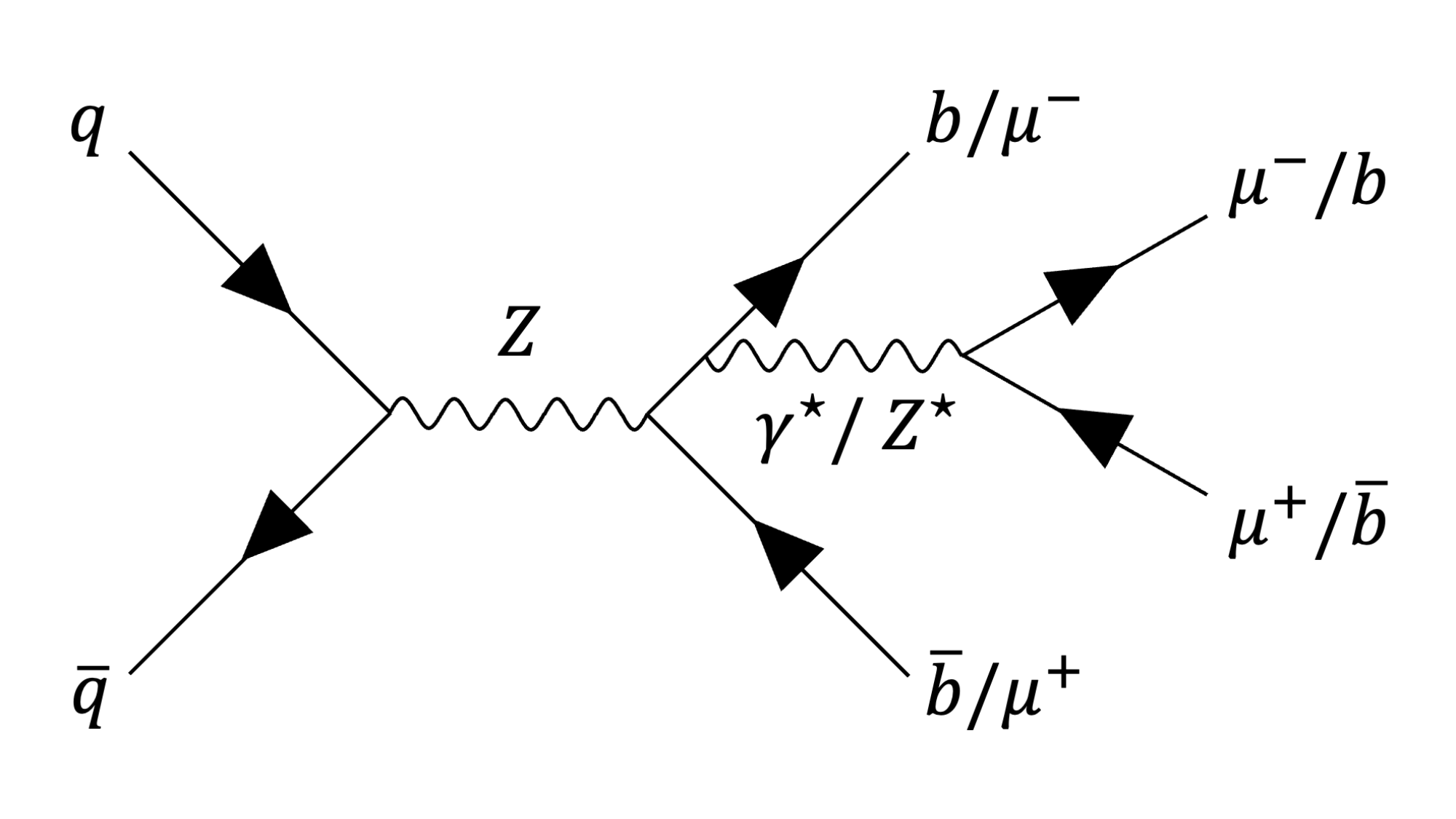}
	\caption{Representative Feynman diagram contributing to the
$\mu^+\mu^-b\bar b$ final state near the $Z$ pole in the Standard Model,
where the four-body final state may arise from final-state radiation
of an off-shell $Z^\ast/\gamma^\ast$.}
	\label{fig:Feynman}
\end{figure}

The diagram shown in FIG.~\ref{fig:Feynman} illustrates the characteristic topology of the $Z \to \mu\mu bb$ decay in the Standard Model, where the four-body final state arises predominantly from final-state radiation of an off-shell electroweak gauge boson. Compared to two-body $Z$ decays, this topology probes a richer kinematic structure and allows access to contact interactions that are absent at leading order in simpler final states. In the SMEFT framework, this process is particularly sensitive to four-fermion operators involving muons and bottom quarks, which generate direct $\mu\mu b b$ interactions at tree level. Such contributions interfere with the SM amplitudes mediated by off-shell $Z^\ast/\gamma^\ast$ exchange, leading to characteristic deviations in both the total rate and differential distributions. As a result, the $Z \to \mu\mu bb$ channel provides a complementary, process-specific probe of lepton–quark operator structures that are only weakly constrained by inclusive $Z$ observables or purely leptonic decay modes.

The analysis is based on leading-order Monte Carlo simulations combined with parton showering and fast detector simulation. Constraints on the selected Wilson coefficients are obtained with an Asimov likelihood-ratio method by varying one operator at a time. Within this setup, the study provides channel-specific constraints on flavor-resolved SMEFT effects in the \ensuremath{\mu^+\mu^-b\bar b} final state. The results also provide a useful starting point for future refinements including higher-order corrections, systematic uncertainties, electroweak interference effects, and multi-operator SMEFT interpretations.

\section{Effective field theory description}

In the absence of direct evidence for new resonances, the effects of physics beyond the Standard Model (SM) can be systematically described using an effective field theory approach \cite{Buchmuller:1985jz,Grzadkowski:2010es,Brivio:2017vri}. Its low-energy effects can be encoded in higher-dimensional operators constructed from SM fields and respecting the SM gauge symmetries.

Within the SMEFT framework, the Lagrangian is organized as an expansion in inverse powers of the new physics scale $\Lambda$,
\begin{equation}
\mathcal{L}_{\mathrm{SMEFT}} =
\mathcal{L}_{\mathrm{SM}}
+ \sum_i \frac{C_i}{\Lambda^2} \mathcal{O}_i
+ \mathcal{O}\left(\frac{1}{\Lambda^4}\right),
\label{eq:smeft_lagrangian}
\end{equation}
where $\mathcal{O}_i$ are dimension-six operators and $C_i$ the corresponding WCs \cite{Falkowski:2015krw,Berthier:2015oma}. In this work, we focus on the effects of dimension-six operators. Both the interference between the SM and dimension-six amplitudes and, in the baseline numerical results, the quadratic contribution from dimension-six amplitudes are considered. The latter is formally of order $1/\Lambda^4$ and should therefore be interpreted with the usual SMEFT truncation caveat.

For a generic event yield, the SMEFT dependence can be written as
\begin{equation}
N(\vec C)
=
N_{\mathrm{SM}}
+
\sum_i
\frac{C_i}{\Lambda^2}
N_i^{\mathrm{int}}
+
\sum_{i,j}
\frac{C_i C_j}{\Lambda^4}
N_{ij}^{\mathrm{quad}},
\label{eq:general_smeft_yield}
\end{equation}
where $N_i^{\mathrm{int}}$ denotes the interference between the SM amplitude 
and the amplitude induced by a single dimension-six operator, while 
$N_{ij}^{\mathrm{quad}}$ denotes the quadratic contribution from dimension-six amplitudes. In the numerical analysis below, one operator is varied at a time, so that Eq.~(\ref{eq:general_smeft_yield}) reduces to a quadratic dependence on a single parameter. The resulting limits are therefore individual expected constraints and should not be interpreted as fully marginalized SMEFT constraints.

The selected \ensuremath{\mu^+\mu^-b\bar b} final state near the 
\ensuremath{Z} pole receives contributions from several classes of dimension-six operators. First, four-fermion operators involving
leptons and bottom quarks generate direct contact interactions contributing
to the four-body final state at tree level \cite{Boughezal:2020uwq}.
Schematically, such operators can be written as $(\bar{\ell} \Gamma \ell)(\bar{q} \Gamma q)$ and induce amplitudes that interfere with the SM contributions mediated by off-shell electroweak bosons.

Second, operators that modify the couplings of the $Z$ boson to fermions lead to indirect contributions to the decay process. After electroweak symmetry breaking, these operators generate shifts in the effective $Z$--lepton and $Z$--quark couplings, which affect both the production and decay of the $Z$ boson \cite{deBlas:2018tjm,Ellis:2018gqa}. These effects propagate into the $Z \to \mu\mu bb$ final state through the modified electroweak vertices. Throughout the event generation and reweighting procedure, the total $Z$-boson width is kept fixed to its SM value. Operators such as $C_{H\ell}^{(1,3)}$ and $C_{Hq}^{(1,3)}$ can in general modify the total and partial $Z$ widths. The expected constraints reported in this work should therefore be interpreted under the fixed-width assumption. Including SMEFT-induced width variations would require a global treatment of electroweak precision observables and is left for future work.

In the numerical analysis, we select a set of six dimension-six operators that contribute to this decay channel and study their impact individually.
The operators are implemented using the \texttt{SMEFTsim\_general\_MwScheme\_UFO} model~\cite{Brivio:2020onw}, which is compatible with \texttt{MadGraph5\_aMC@NLO} \cite{Alwall:2014hca} and allows event generation and reweighting at leading order in a consistent electroweak input scheme.

In the baseline numerical results, both interference and quadratic dimension-six terms are retained in the event yield. This convention follows the structure of the squared matrix element used in the reweighting procedure and is particularly relevant for operators with suppressed interference. The associated EFT truncation caveat is discussed above and is taken into account when interpreting the quoted constraints.

The validity of the SMEFT expansion relies on the assumption that the typical energy scale of the process remains below the cutoff scale $\Lambda$
\cite{Elias-Miro:2013mua,Falkowski:2015krw}. Within the selected \ensuremath{Z}-pole region, the analysis is dominated by moderate 
kinematic scales set by the reconstructed invariant mass and object transverse momenta, which supports the use of the SMEFT description in the phase space considered. A simultaneous multi-operator fit is beyond the scope of the present exploratory study. Since correlations among operators can weaken individual constraints and may lead to approximate flat directions, the bounds quoted below represent individual one-operator-at-a-time expected constraints. A global treatment including correlations with other electroweak and flavor observables would be required for a complete SMEFT interpretation.

\section{Event generation and simulation}

 All processes are generated at leading order using \texttt{MadGraph5\_aMC@NLO}, followed by parton showering and hadronization with \texttt{Pythia8} \cite{Sjostrand:2014zea}. Detector-level effects are simulated using \texttt{Delphes} \cite{deFavereau:2013fsa} with a CMS-like detector configuration. Unless stated otherwise, proton--proton collisions at a center-of-mass energy of $\sqrt{s} = 13~\mathrm{TeV}$ are assumed throughout this analysis. All numerical results presented below should be understood as leading-order constraints within the simulation setup described here. Higher-order QCD and electroweak corrections, as well as scale and PDF uncertainties, are not included in the baseline simulation. For several of the processes considered here, next-to-leading-order predictions can be obtained with public Monte Carlo tools and may modify both the normalization and the shapes of kinematic distributions. Their inclusion is left for future precision studies.

The primary final state considered in this work is $pp\to\mu^+\mu^-b\bar b$ in the reconstructed $Z$-pole region. For brevity, we refer to this topology as the $Z\to\mu\mu b\bar b$ channel. At the amplitude level, however, the same final state may receive contributions from off-shell electroweak bosons and photon-mediated diagrams. The present analysis should therefore be understood as a fiducial sensitivity study of the $\mu^+\mu^-b\bar b$ final state around the $Z$ resonance rather than a fully gauge-invariant separation of individual diagram classes.

Signal events are generated at parton level using \texttt{MadGraph5\_aMC@NLO}. Basic generator-level cuts are applied to ensure numerical stability and reflect the kinematic acceptance of the detector: $p_T^\ell>10~\mathrm{GeV}$, $|\eta^\ell|<2.5$ for charged leptons, and $p_T^q>10~\mathrm{GeV}$, $|\eta^q|<2.5$ for quarks. The resulting leading-order cross section for the selected signal topology is $\sigma = 0.750~\mathrm{fb}$ at generator level before the full detector-level selection.

\begin{figure}[H]
	\centering

	\subfloat[Invariant mass distributions of $\mu\mu$, $bb$, and $\mu\mu bb$.]{
		\includegraphics[width=0.85\columnwidth]{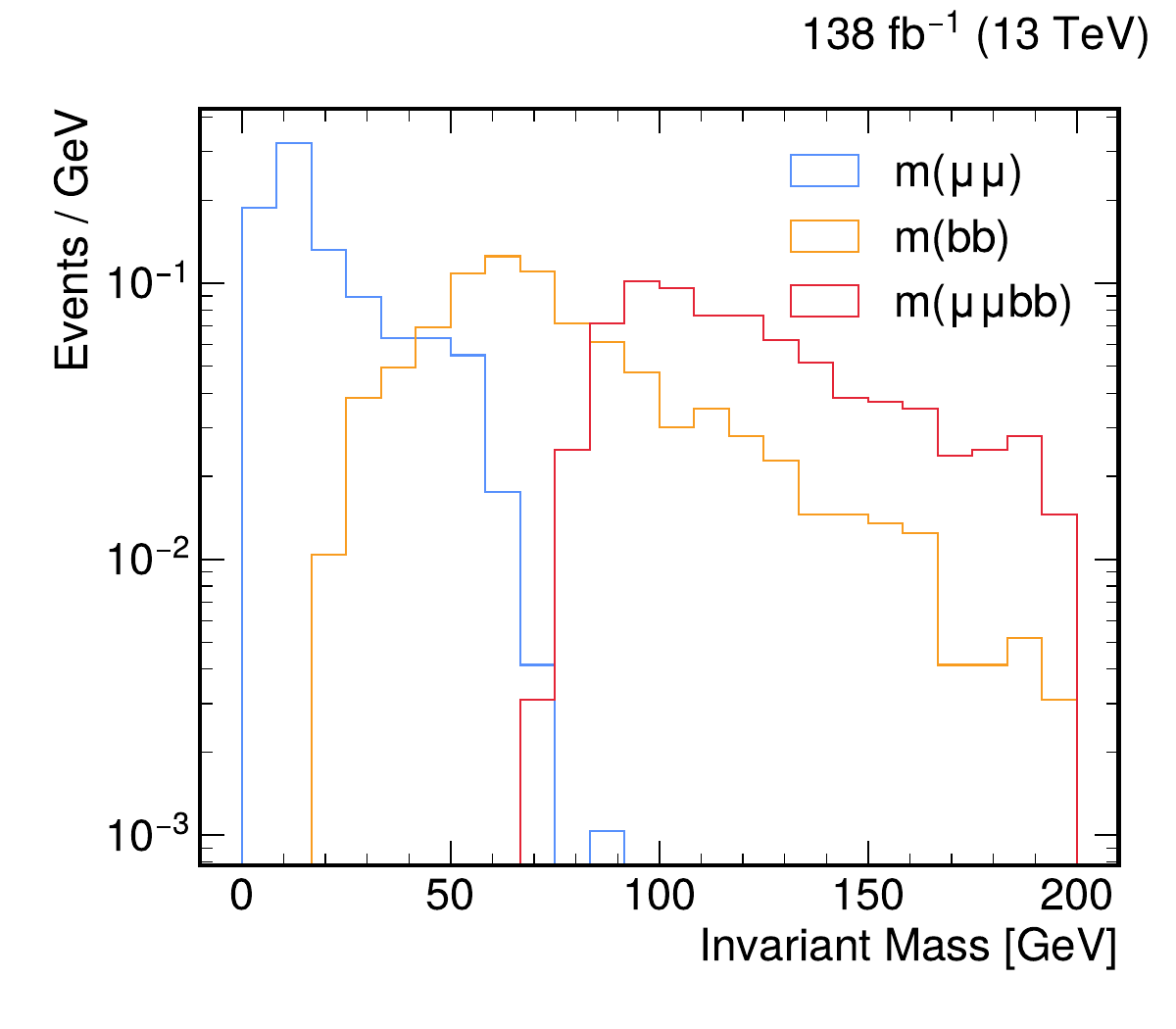}
		\label{fig:m2m2b}
	}

	\vspace{0.5cm}

	\subfloat[Transverse momentum distributions of muons and jets.]{
		\includegraphics[width=0.85\columnwidth]{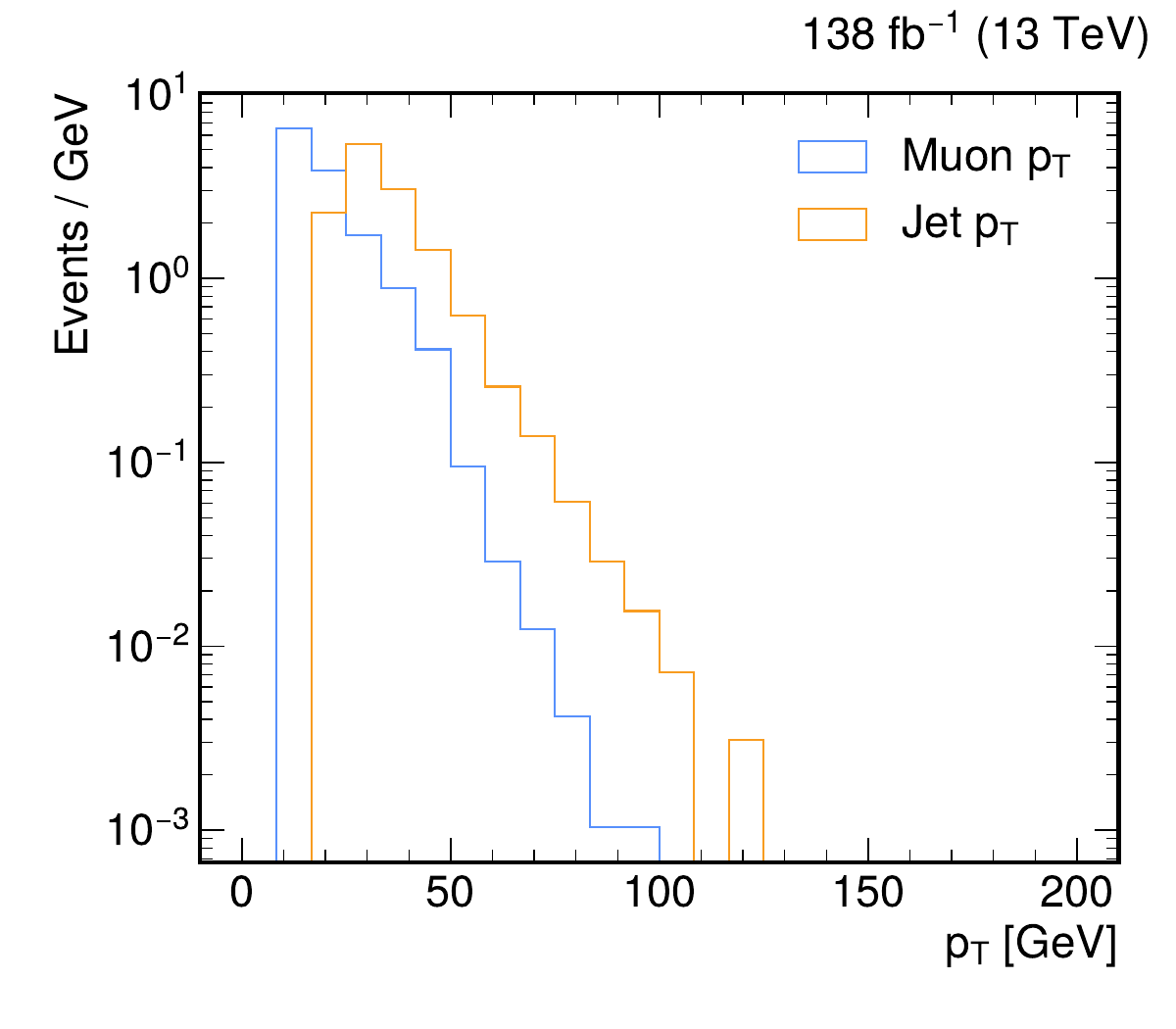}
		\label{fig:pt2m2b}
	}

	\caption{Detector-level kinematic distributions for the selected 
\ensuremath{\mu^+\mu^-b\bar b} final-state topology obtained from 
\texttt{Delphes} simulation.}

	\label{fig:2m2b}
\end{figure}

The reconstructed kinematic distributions shown in Fig.~\ref{fig:2m2b} provide an important validation of the signal modeling after detector effects are taken into account. In Fig.~\ref{fig:m2m2b}, the invariant mass distribution of the \ensuremath{b\bar b} system is considerably 
broader than that of the dimuon system, as expected from jet energy resolution effects and QCD radiation. Nevertheless, the combined four-body invariant mass $m_{\mu\mu bb}$ displays a clear enhancement around $m_Z$, indicating that the signal topology can be reliably reconstructed at detector level despite the presence of hadronic final states.

Fig.~\ref{fig:pt2m2b} shows the transverse momentum spectra of the reconstructed muons and jets. The muon $p_T$ distribution is relatively soft, characteristic of a $Z$-boson decay, while the jet $p_T$ spectrum features a longer tail driven by QCD dynamics. The typical energy scales involved remain well below the assumed SMEFT cutoff, supporting the consistency of the SMEFT interpretation within the selected phase space.

The dominant background arises from top-quark pair production $pp \to t\bar{t}$, with both top quarks decaying semileptonically, $t \to b \mu^+ \nu_\mu$ and $\bar{t} \to \bar{b} \mu^- \bar{\nu}_\mu$ \cite{Alwall:2014hca,Sjostrand:2014zea}. Due to its large production cross section, this process remains a significant background even after event selection.

Additional backgrounds include diboson production, such as $pp \to ZZ$, $Z \to \mu \mu$, $Z \to bb$ as well as associated production of a Z boson with a Higgs boson: $pp \to ZH$, $Z \to \mu^+ \mu^-$, $H \to b\bar{b}$. Although their cross sections are smaller than that of $t\bar{t}$ production, these processes closely resemble the signal topology and are therefore included explicitly \cite{deFavereau:2013fsa,Alwall:2014hca}.

The Drell--Yan production of a $Z$ boson in association with jets: $pp \to Z + \text{jets}$, constitutes another important background due to its very large cross section. Events with light-flavor jets can enter the signal region through jet misidentification, while events with heavy-flavor jets provide an irreducible background. To model this background accurately, samples with up to two additional jets at matrix-element level are generated and merged with the parton shower using the MLM matching scheme \cite{Mangano:2006rw,Alwall:2007fs}.

Subleading backgrounds from electroweak processes involving photons are also considered, including $Z\gamma$, $\gamma\gamma$, and mixed $\gamma Z$ production, where the photon converts into a fermion pair. We note that electroweak amplitudes leading to the same $\mu^+\mu^-b\bar b$ final state can in principle interfere. In the present exploratory analysis, the samples are organized according to their dominant physical topology in order to 
estimate their relative impact after selection. A fully gauge-invariant treatment of the irreducible electroweak $\mu^+\mu^-b\bar b$ contribution, including all interferences among $Z^\ast$- and $\gamma^\ast$-mediated diagrams, is beyond the scope of this work. This point is taken into account by interpreting the numerical results as channel-specific constraints within the event-generation setup described above.

A summary of all signal and background processes and their corresponding
leading-order cross sections is given in TABLE~\ref{tab:xsec}.

\begin{table}[H]
\centering
\renewcommand{\arraystretch}{1.5} % 调整行高，1.5 倍
\begin{tabular}{l c}
\hline\hline
Process & Cross section [pb] \\
\hline
$pp \to \mu^+\mu^- b\bar{b}$, $Z$-pole topology & $7.50 \times 10^{-4}$ \\
$pp \to t\bar{t}$, $t \to b \mu^+ \nu_\mu, \bar{t} \to \bar{b} \mu^- \bar{\nu}_\mu$ & $4.38 $ \\
$pp \to ZZ$, $Z \to \mu^+ \mu^-$, $Z \to b\bar{b}$ & $5.54 \times 10^{-2} $ \\
$pp \to Z + \text{jets}$ $(0,1,2)$ & $9.33 \times 10^{2} $ \\
$pp \to ZH$, $Z \to \mu^+ \mu^-$, $H \to b\bar{b}$ & $1.03 \times 10^{-2}$ \\
$pp \to Z\gamma$, $\gamma \to b\bar{b}$ & $5.54 \times 10^{-3} $ \\
$pp \to \gamma\gamma$, $\gamma \to \mu^+ \mu^-,\ \gamma \to b\bar{b}$ & $6.70 \times 10^{-5} $ \\
\hline\hline
\end{tabular}
\caption{Leading-order cross sections for the signal and background processes
considered in this analysis. The values are obtained at generator level before the full detector-level event selection and are intended to summarize the relative normalization of the simulated samples. The merged $Z$+jets sample, combining matrix-element calculations with up to two additional jets, is used in the final analysis.}

\label{tab:xsec}
\end{table}

All generated parton-level events are passed to \texttt{Pythia8} for parton
showering and hadronization. For the $Z+\text{jets}$ background, matrix-element calculations with up to two additional jets are merged with the parton shower using the MLM matching scheme \cite{Mangano:2006rw,Alwall:2007fs}. This procedure avoids double counting between matrix-element and parton-shower emissions and provides a more reliable description of multi-jet final states.

Detector effects are simulated with \texttt{Delphes}, using a CMS-like detector configuration. Jets are identified as $b$-jets using the built-in $b$-tagging algorithm, and a fixed working point corresponding to a discriminator threshold of $\text{BTag} > 0.5$ is applied. This working point yields an average $b$-tagging efficiency of approximately $70\%$ for true $b$-jets, with mis-tag rates of about $10\%$ for charm jets and $1\%$ for light-flavor jets, consistent with typical CMS performance~\cite{collaboration_2013}. The requirement of two $b$-tagged jets significantly suppresses backgrounds from light-flavor and charm jet production, in particular from $Z+$jets events.

\section{Event Reconstruction and Selection Strategy}

Events are reconstructed using the detector-level objects provided by \texttt{Delphes}. Muon candidates are reconstructed within the pseudorapidity range $|\eta| < 2.5$, corresponding to the coverage of the central tracking system and muon chambers in LHC experiments \cite{Chatrchyan:2008aa, Aad:2010xm}. Jets are reconstructed using the anti-$k_T$ algorithm with a radius parameter $R = 0.4$ \cite{Cacciari:2008gp}.

Bottom-quark jets are identified using the $b$-tagging information stored in the \texttt{Delphes} output \cite{deFavereau:2013fsa, collaboration_2013}. A jet is considered a $b$-jet if its $b$-tag discriminator exceeds a threshold of $0.5$, corresponding to a typical medium working point used in CMS analyses.

To emulate the trigger requirements of LHC experiments, events are required to satisfy at least one of the muon-based trigger conditions listed in
TABLE~\ref{tab:trigger}. These triggers are motivated by standard CMS single- and double-muon triggers used in Run~\text{II} analyses \cite{CMS:2017wtu}. The single-muon trigger ensures high efficiency for events containing a high-$p_T$ muon, while the di-muon trigger retains sensitivity to events in which both muons have moderate transverse momenta. Only isolated muons are considered for trigger selection to suppress backgrounds from heavy-flavor decays and hadronic activity.

\begin{table}[H]
\centering
\renewcommand{\arraystretch}{1.5} % 调整行高，1.5 倍
\begin{tabular}{l c}
\hline\hline
Trigger requirement & Threshold [GeV] \\
\hline
Single-muon trigger & $p_T(\mu) > 25$ \\
Double-muon trigger (leading) & $p_T(\mu_1) > 18$ \\
Double-muon trigger (subleading) & $p_T(\mu_2) > 7$ \\
\hline\hline
\end{tabular}
\caption{Muon trigger requirements used in the analysis.
Only isolated muons are considered for trigger selection.}
\label{tab:trigger}
\end{table}

Events passing the trigger requirements are subjected to a set of preselection criteria designed to suppress reducible backgrounds while maintaining high signal efficiency. The preselection cuts are summarized in TABLE~\ref{tab:preselection}.

At least two muon candidates are required per event. The subleading muon is required to satisfy $p_T > 10~\mathrm{GeV}$ to ensure reliable reconstruction and identification efficiency. Muon isolation is imposed by requiring a minimum separation $\Delta R(\mu, \text{jet}) > 0.4$ between each selected muon and any reconstructed jet \cite{Cacciari:2008gp}. This reduces backgrounds from semileptonic heavy-flavor decays.

To suppress backgrounds containing genuine missing transverse energy, such as top-quark pair production, events are required to have
$\mathrm{MET} < 30~\mathrm{GeV}$. This exploits the absence of prompt neutrinos in the selected \ensuremath{\mu^+\mu^-b\bar b} signal topology.

Events are required to contain at least two reconstructed jets with $p_T > 25~\mathrm{GeV}$, among which at least two must be identified as $b$-jets. This reflects the expected presence of a $b\bar{b}$ pair in the signal final state and significantly suppresses backgrounds from $Z+$ light-jets production.

Angular separation requirements are imposed to ensure well-resolved final-state objects. In particular, the separation between the two selected $b$-jets must satisfy $\Delta R(b,b) > 0.4$, and the separation between the two leading muons must satisfy $\Delta R(\mu,\mu) > 0.4$. Additionally, all muon--$b$-jet pairs must satisfy $\Delta R(\mu,b) > 0.4$ \cite{Cacciari:2008gp}. These requirements reduce contamination from collinear final-state radiation and overlapping reconstruction effects.

\begin{table}[H]
\centering
\renewcommand{\arraystretch}{1.5} % 调整行高，1.5 倍
\begin{tabular}{l c}
\hline\hline
Selection & Requirement \\
\hline
Number of muons & $\geq 2$ \\
Muon $p_T$ (subleading) & $> 10~\mathrm{GeV}$ \\
Muon isolation & $\Delta R(\mu,\text{jet}) > 0.4$ \\
Missing transverse energy & $\mathrm{MET} < 30~\mathrm{GeV}$ \\
Number of jets & $\geq 2$ \\
Jet $p_T$ & $> 25~\mathrm{GeV}$ \\
$b$-tagging & $\geq 2$ $b$-jets \\
$\Delta R(b,b)$ & $> 0.4$ \\
$\Delta R(\mu,\mu)$ & $> 0.4$ \\
$\Delta R(\mu,b)$ & $> 0.4$ \\
\hline\hline
\end{tabular}
\caption{Summary of event selection criteria applied in the analysis.}
\label{tab:preselection}
\end{table}

The selection criteria are designed to preserve the characteristic kinematics of the $Z \to \mu\mu bb$ decay while suppressing backgrounds with genuine missing energy or misidentified heavy-flavor jets. For events passing all selection requirements, the invariant mass of the $\mu \mu bb$ system is reconstructed using the four-momenta of the 
two leading isolated muons and the two selected $b$-jets. The reconstructed invariant mass $m_{\mu\mu bb}$ serves as the primary observable for signal extraction and the subsequent SMEFT analysis \cite{Boughezal:2020uwq}.

The signal region is defined as $80~\mathrm{GeV} < m_{\mu\mu bb} < 115~\mathrm{GeV}$, centered around the $Z$ boson mass. This window retains a large fraction of the selected signal topology while reducing non-resonant backgrounds. It provides a simple and experimentally motivated signal region for the subsequent SMEFT constraint extraction.

\begin{figure}[H]
    \centering

    \subfloat[Invariant mass distributions in the $(70,250)$ ~GeV range, full spectrum with signal scaled by 100 for visibility.]{
        \includegraphics[width=0.9\columnwidth]{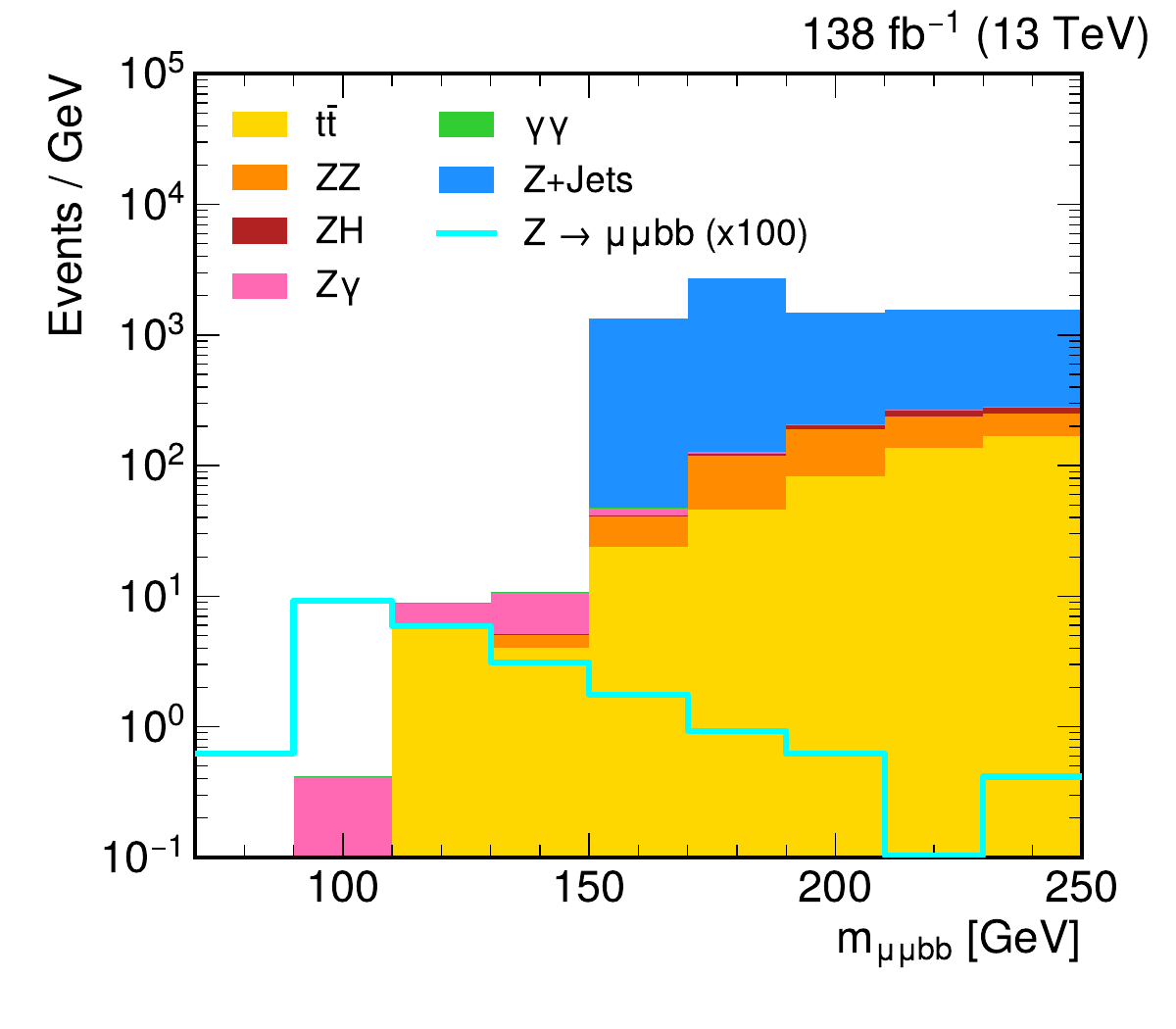}
        \label{fig:mumubb_70_250}
    }

    \vspace{0.5cm}

    \subfloat[Invariant mass distributions in the $(80,115)$ ~GeV range, focusing on the signal region.]{
        \includegraphics[width=0.875\columnwidth]{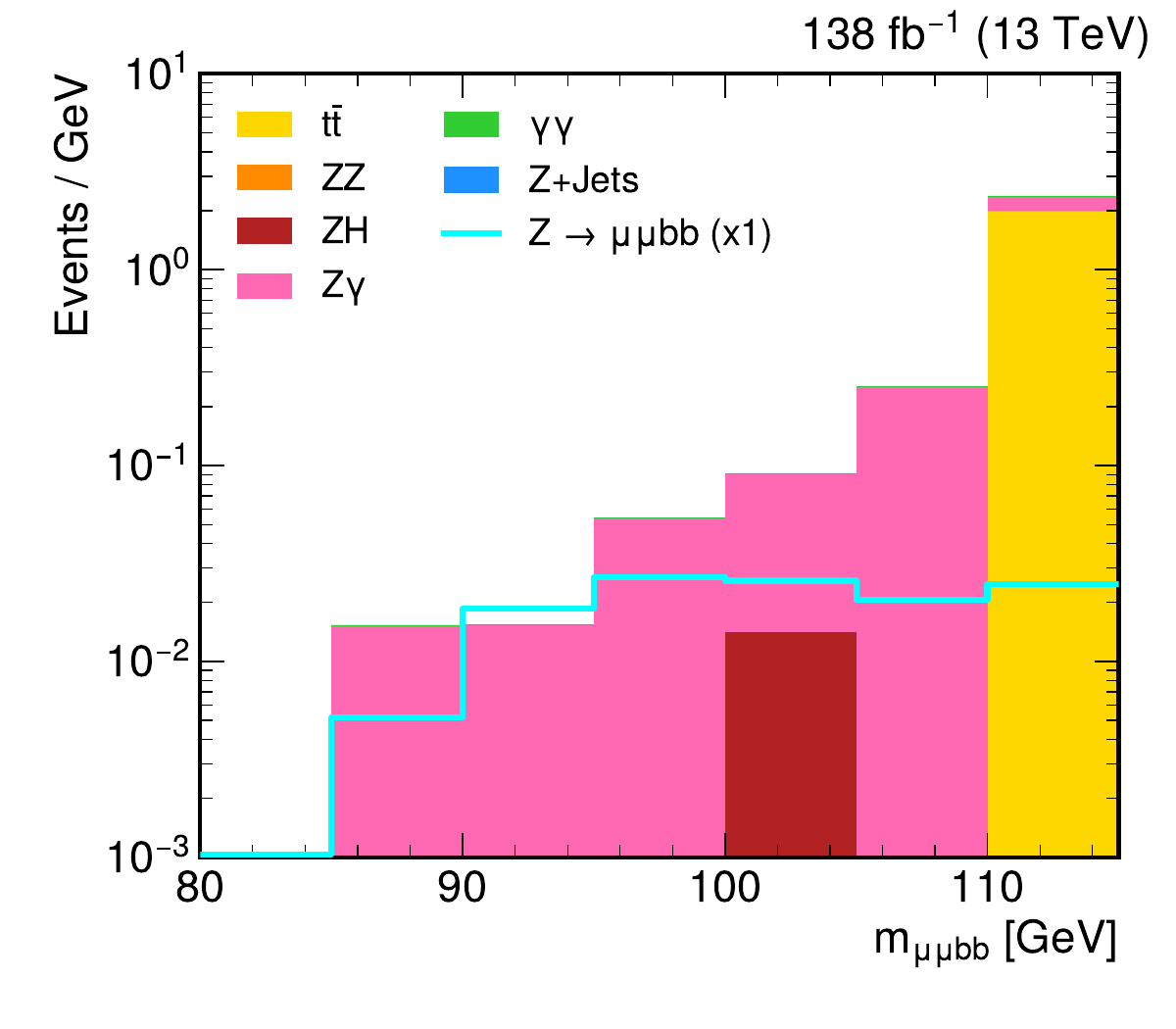}
        \label{fig:mumubb_80_115}
    }

    \caption{Detector-level invariant mass distributions for $\mu\mu bb$ from various processes, obtained from \texttt{Delphes} simulation. The first panel shows the full mass spectrum with the signal scaled for better visibility, while the second panel zooms into the signal region.}
    \label{fig:mumubb_all}
\end{figure}

FIG.~\ref{fig:mumubb_all} compares the reconstructed invariant mass distributions of the $\mu\mu bb$ system for the signal and the dominant background processes after the full event selection. In FIG.~\ref{fig:mumubb_70_250}, the distribution is shown over a wide mass range from 70 to 250~GeV, with the signal contribution scaled by a factor of 100 for visibility. Over this range, the spectrum is dominated by the $Z$+jets and $t\bar{t}$ backgrounds, reflecting their large production cross sections.

FIG.~\ref{fig:mumubb_80_115} focuses on the signal region defined by 
$80 < m_{\mu\mu bb} < 115$~GeV. In this window, the relative contribution of 
the selected signal topology is enhanced, while the $t\bar t$ background is 
reduced by the missing transverse energy requirement. The remaining electroweak and associated-production backgrounds, including $Z\gamma$ and $ZH$ production, exhibit similar kinematic features but are subleading in rate after the applied selection. The invariant mass $m_{\mu\mu bb}$ is therefore used as the main observable defining the signal region for the subsequent reweighting and statistics-only constraint extraction.

\section{SMEFT Reweighting and Constraint Extraction}

The impact of dimension-six operators on the selected $\mu^+\mu^-b\bar b$ final-state topology is evaluated within the SMEFT framework using an event reweighting technique \cite{Conte:2012fm, Alloul:2013bka}. In this approach, the dependence of the selected event yield on each WC is obtained by reweighting Monte Carlo events. The total event weight includes the SM 
contribution as well as linear and quadratic corrections from dimension-six 
operators, allowing an efficient exploration of the WC parameter space without regenerating independent samples for every parameter point.

SMEFT effects are incorporated using the \texttt{SMEFTsim\_general\_MwScheme\_UFO} model from the SMEFTsim~3.0 framework \cite{Brivio:2020onw}, which provides a gauge-invariant implementation of the relevant dimension-six operators in the $M_W$ input scheme. The $M_W$ scheme is adopted to consistently account for electroweak input-parameter shifts induced by dimension-six operators. For each generated event, a set of precomputed weights corresponding to different 
values of $\theta$ is stored at generator level. After parton showering, 
hadronization, detector simulation, and event selection, these weights are used to parametrize the SMEFT dependence of the selected yield.

After applying the full event selection, the expected number of signal
events is expressed as a quadratic function of the WCs parameter $\theta \equiv C_i / \Lambda^2$,
\begin{equation}
N(\theta) = c + b\,\theta + a\,\theta^2 ,
\label{eq:Ntheta}
\end{equation}
where $c$ denotes the Standard Model (SM) expectation, $b$ corresponds to
the interference between the SM and dimension-six amplitudes, and $a$
encodes the pure quadratic SMEFT contribution \cite{Boughezal:2020uwq}.
In principle, the reweighting information allows the SM contribution, the 
SM--dimension-six interference term, and the quadratic dimension-six contribution to be extracted separately. In the present analysis, the coefficients $a$, $b$, and $c$ are determined by fitting the reweighted event yields at several benchmark values of $\theta$. This provides a simple numerical implementation of the same quadratic dependence and is used as a consistency check of the reweighting behavior after event selection. The fitted parametrization is used only in the parameter region relevant for the quoted confidence intervals.

Constraints on the WCs are derived using a likelihood-based
statistical approach \cite{Cowan:2010js}. An Asimov dataset is constructed
by assuming the observed number of events to be equal to the SM prediction,
$N_{\mathrm{obs}} = N_{\mathrm{SM}} \equiv c$.
The Poisson likelihood function is then given by
\begin{equation}
\mathcal{L}(\theta)
= \frac{[N(\theta)]^{N_{\mathrm{obs}}} e^{-N(\theta)}}{N_{\mathrm{obs}}!}.
\end{equation}

The test statistic is defined as the likelihood ratio
\begin{equation}
q(\theta)
= -2 \ln \frac{\mathcal{L}(\theta)}{\mathcal{L}(0)}
= 2 \left[
N(\theta)\ln\frac{N(\theta)}{N_{\mathrm{SM}}}
+ N_{\mathrm{SM}} - N(\theta)
\right],
\label{eq:qtheta}
\end{equation}
which is compared with a $\chi^2$ distribution with one degree of freedom under the large-sample approximation of Wilks' theorem \cite{Wilks:1938dza}.
The 95\% confidence level (C.L.) interval on $\theta$ is obtained by requiring $q(\theta) = 3.84$. No nuisance parameters are included in the baseline likelihood. The resulting intervals therefore correspond to statistics-only expected constraints within the baseline likelihood setup.

This procedure is applied independently to each dimension-six operator,
with all other WCs set to zero, providing channel-specific constraints on possible SMEFT effects in the \ensuremath{\mu^+\mu^-b\bar b} final state. In this analysis, a flavor-specific setup is adopted, in which the WCs are defined for individual fermion generations. In particular, operators involving second-generation leptons ($\ell_2$) and third-generation quarks ($q_3$) are considered explicitly. This choice provides channel-specific constraints on flavor-resolved SMEFT effects in the \ensuremath{\mu^+\mu^-b\bar b} final state, beyond the flavor-universal assumptions commonly employed in global SMEFT fits.

\begin{figure}[H]
    \centering

    \subfloat[$C_{H \ell}^{(1),22}$]{\includegraphics[width=.5\columnwidth]{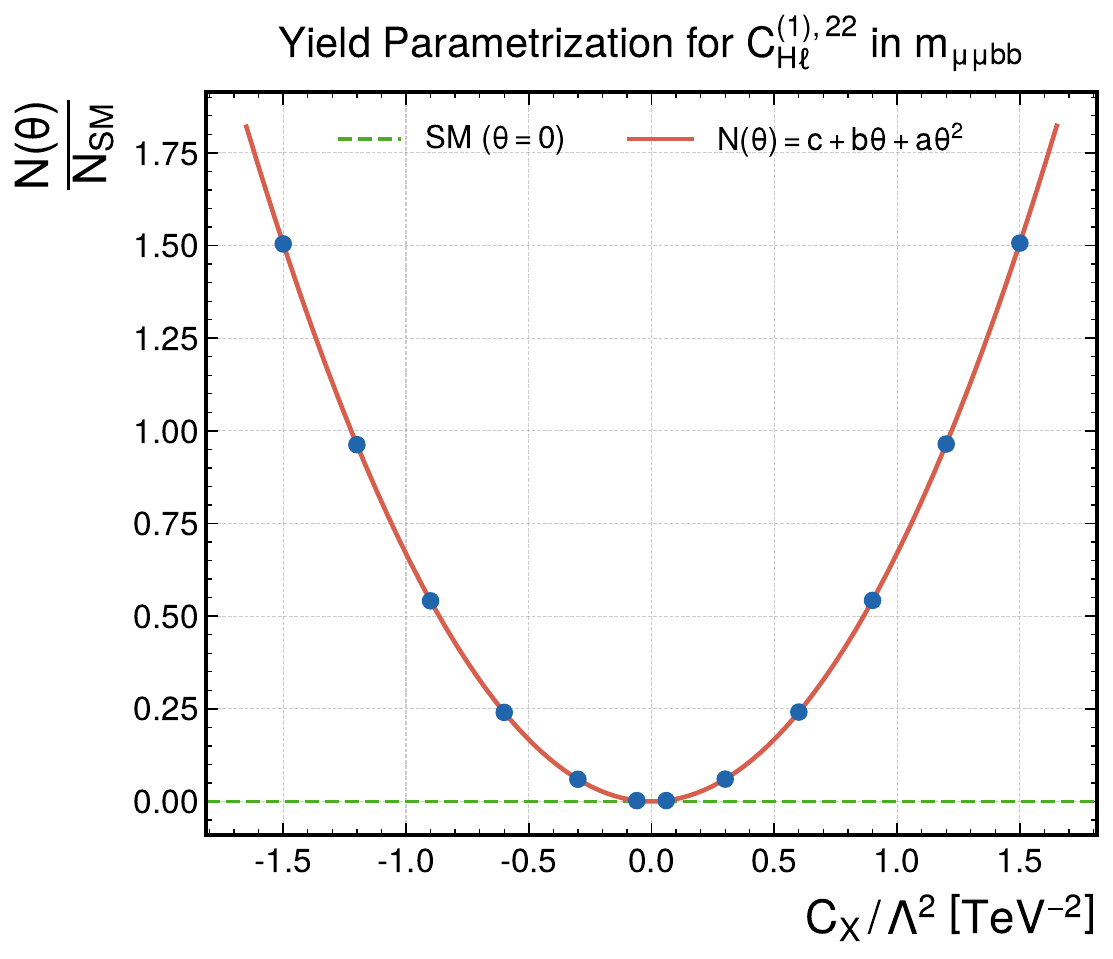}}
    \subfloat[$C_{H \ell}^{(3),22}$]{\includegraphics[width=.5\columnwidth]{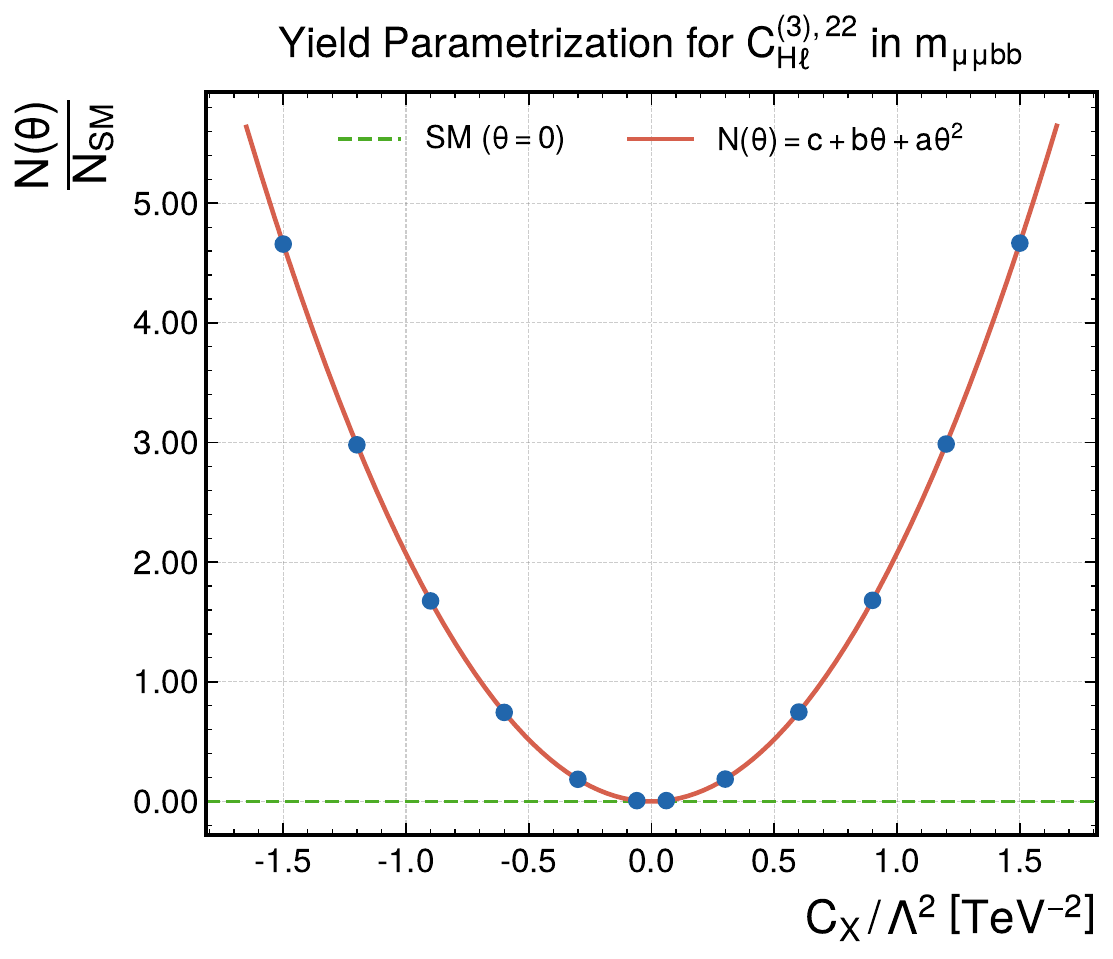}}

    \subfloat[$C_{H q}^{(1),33}$]{\includegraphics[width=.5\columnwidth]{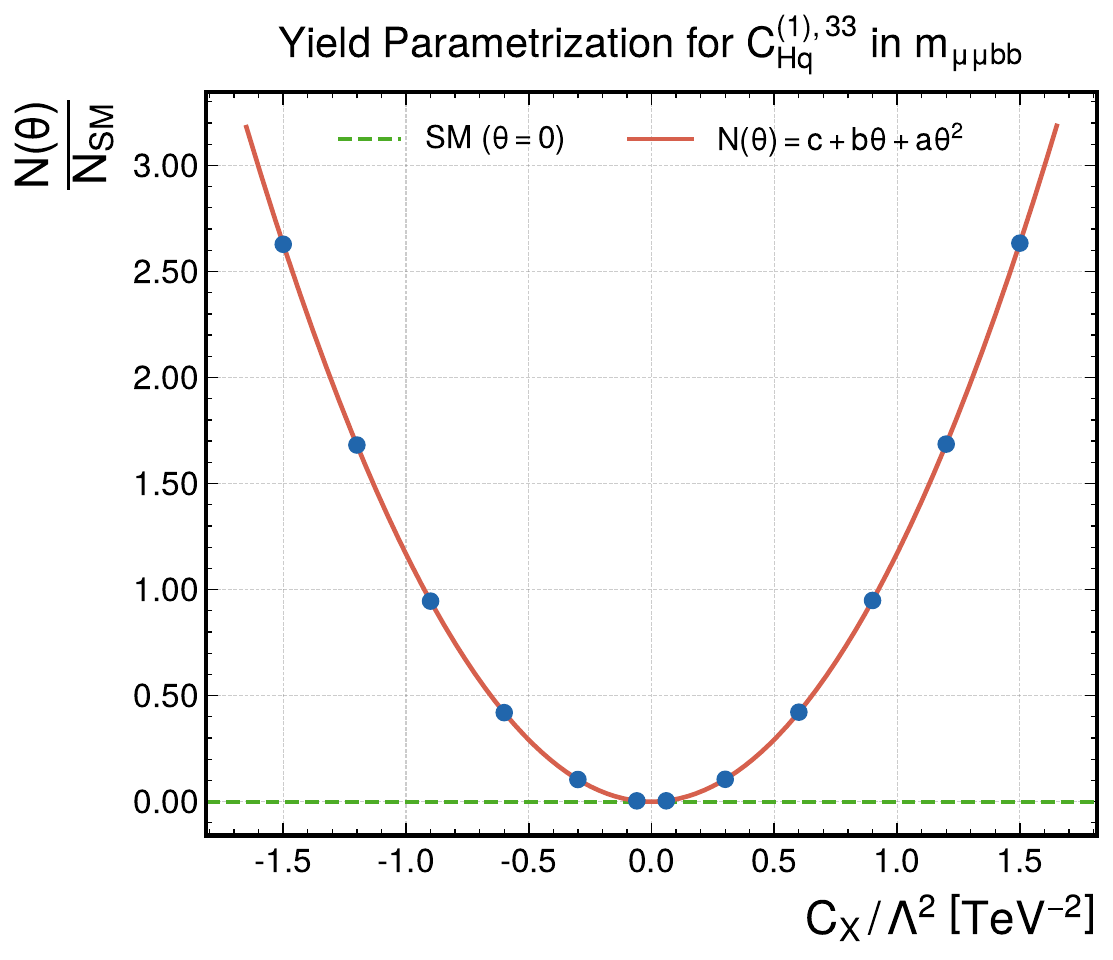}}
    \subfloat[$C_{H q}^{(3),33}$]{\includegraphics[width=.5\columnwidth]{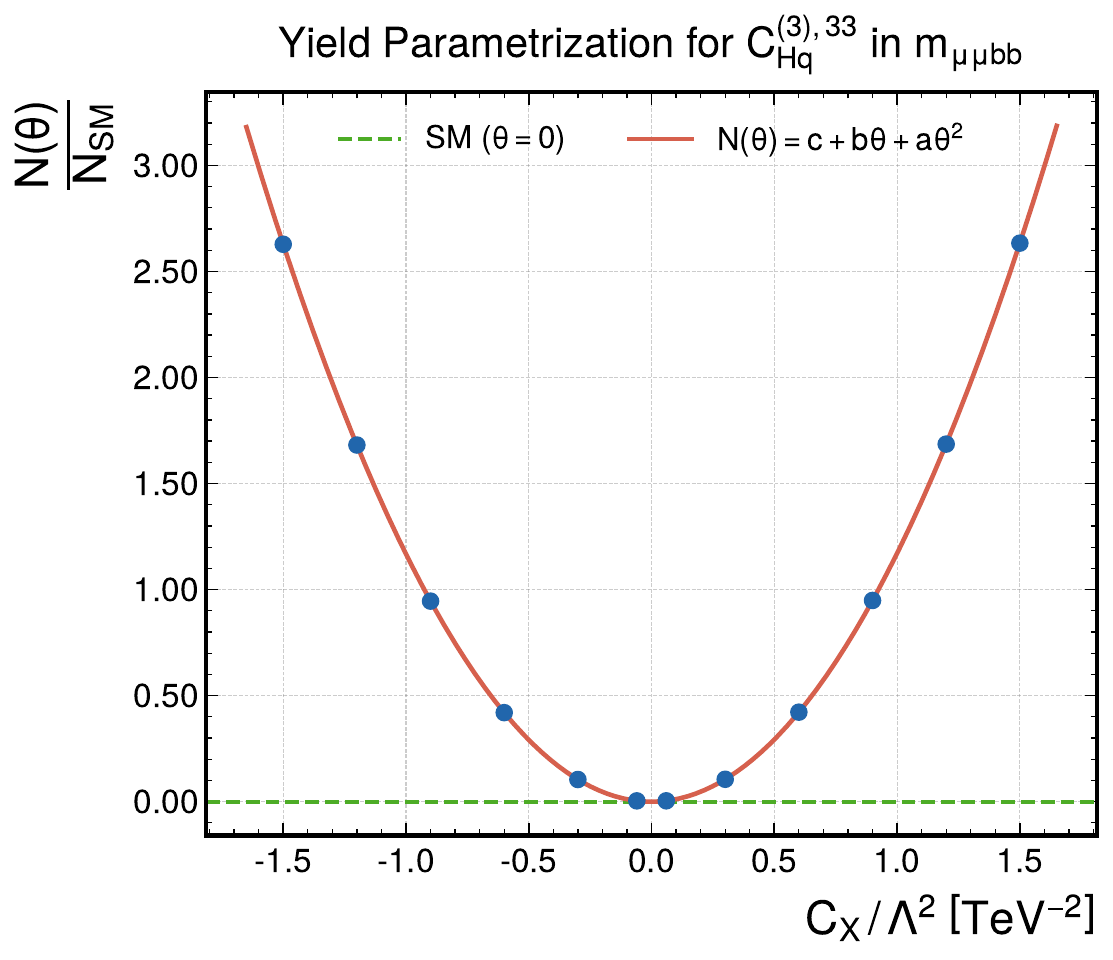}}

    \subfloat[$C_{\ell q}^{(1),2233}$]{\includegraphics[width=.5\columnwidth]{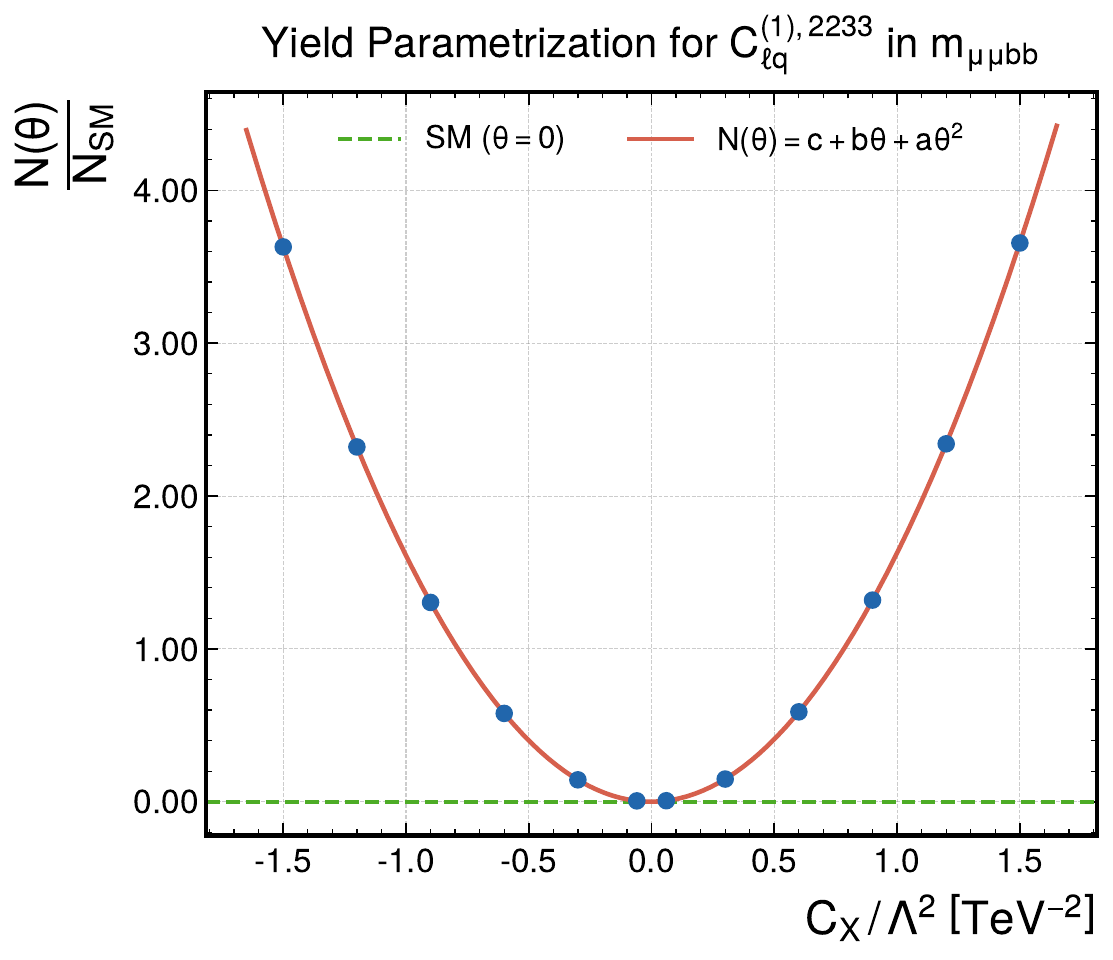}}
    \subfloat[$C_{\ell q}^{(3),2233}$]{\includegraphics[width=.5\columnwidth]{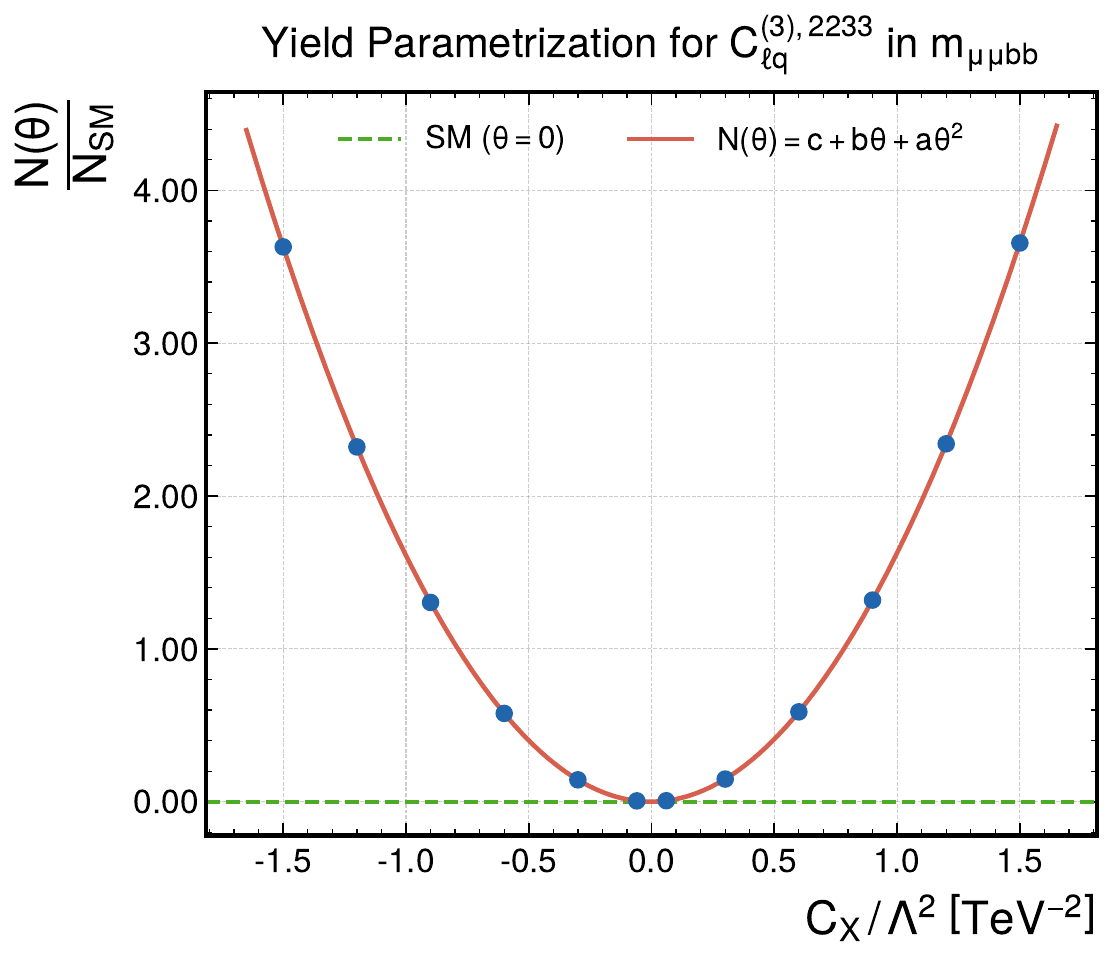}}

    \caption{Dependence of the selected signal-region yield on individual Wilson coefficients in the parameter range relevant for the expected 95\% C.L. intervals. The curves are obtained from the quadratic parametrization $N(\theta)=c+b\theta+a\theta^2$ using SMEFT reweighted samples. Only one coefficient is varied at a time, while all other coefficients are set to zero. The results include both interference and quadratic dimension-six contributions.}

    \label{fig:eft_fit}
\end{figure}

FIG.~\ref{fig:eft_fit} shows the dependence of the expected event yield in the signal region on the WCs of the six dimension-six operators considered in this analysis. For each operator, the ratio of the SMEFT prediction to the SM expectation is fitted with a quadratic function, reflecting the structure of the squared matrix element, which includes both interference and pure SMEFT contributions.

The observed quadratic behavior is consistent with the expected structure of the squared matrix element when one Wilson coefficient is varied at a time. It also illustrates the relative importance of linear and quadratic terms for different operator classes. Operators modifying the $Z$--fermion couplings exhibit sizable interference with the SM amplitude, leading to asymmetric constraints around the SM point. In contrast, certain four-fermion operators show a more symmetric dependence, indicating a dominant contribution from the quadratic term. This behavior highlights the particular sensitivity of this channel to four-fermion operators involving bottom quarks.

The resulting expected 95\% C.L. intervals on the WCs obtained from this procedure are summarized in Table~\ref{tab:eft_theta95}. These intervals are statistics-only, one-operator-at-a-time expected constraints obtained at leading order.

A direct numerical comparison with existing global SMEFT fits should be made with care. Global analyses typically combine electroweak precision data, Higgs measurements, diboson production, top-quark observables, and differential LHC measurements. They may also adopt different flavor assumptions, operator subsets, input schemes, and SMEFT truncation prescriptions. In particular, bounds may differ significantly depending on whether they are obtained one operator at a time or after marginalization over a larger parameter space, and whether quadratic dimension-six terms are retained.

For this reason, the results in Table~\ref{tab:eft_theta95} should be viewed as channel-specific expected constraints rather than direct replacements for global SMEFT constraints. Their main value is to illustrate that the 
$\mu^+\mu^-b\bar b$ final state has potential sensitivity to flavor-resolved 
operators involving second-generation leptons and third-generation quarks, especially four-fermion structures that are not always explicitly resolved in flavor-universal fits. The HL-LHC luminosity scenario illustrates the statistical improvement expected at larger datasets, while a full high-luminosity interpretation would also require a dedicated treatment of systematic uncertainties.

\begin{table}[H]
\centering
\renewcommand{\arraystretch}{1.5} % 调整行高，1.5 倍
\begin{tabular}{l @{\hskip 8mm} c @{\hskip 8mm} c} % 在两列之间增加 10mm 间距
\hline\hline
$C$ & 138 fb$^{-1}$ & 3000 fb$^{-1}$ \\
\hline
$C_{\ell q}^{(1),2233}$ & $[-0.023,\;0.014]$ & $[-0.005,\;0.003]$ \\
$C_{\ell q}^{(3),2233}$ & $[-0.023,\;0.014]$ & $[-0.005,\;0.003]$ \\
$C_{H q}^{(1),33}$      & $[-0.029,\;0.026]$ & $[-0.006,\;0.006]$ \\
$C_{H q}^{(3),33}$      & $[-0.030,\;0.026]$ & $[-0.006,\;0.006]$ \\
$C_{H \ell}^{(1),22}$   & $[-0.025,\;0.009]$ & $[-0.005,\;0.002]$ \\
$C_{H \ell}^{(3),22}$   & $[-0.028,\;0.025]$ & $[-0.006,\;0.005]$ \\
\hline\hline
\end{tabular}
\caption{95\% C.L. allowed regions for $C/ \Lambda^2$ [TeV$^{-2}$]. The results are obtained at leading order using the fixed-width approximation and include both interference and quadratic dimension-six contributions.}
\label{tab:eft_theta95}
\end{table}

\subsection{Uncertainty treatment}

The likelihood used in the baseline results includes only the Poisson statistical uncertainty of the expected event yield. Experimental systematic uncertainties, such as luminosity, muon reconstruction efficiency, $b$-tagging efficiency, jet energy scale, and background normalization uncertainties, are not included. The resulting intervals therefore represent statistics-only expected constraints in the baseline setup.

A complete treatment of systematic uncertainties would require nuisance parameters constrained by auxiliary measurements. Schematically, for each bin or signal region one may write
\begin{equation}
N_k(\vec C,\vec\eta)
=
N_{k,\mathrm{sig}}(\vec C)(1+\eta_s)
+
\sum_b N_{k,b}(1+\eta_b),
\end{equation}
with Gaussian constraints on the nuisance parameters,
\begin{equation}
\mathcal{L}(\vec C,\vec\eta)
=
\prod_k
\mathrm{Pois}
\left(
n_k \mid N_k(\vec C,\vec\eta)
\right)
\prod_j
\exp
\left(
-\frac{\eta_j^2}{2\sigma_j^2}
\right).
\end{equation}
Including such effects would generally weaken the projected bounds, especially for the HL-LHC luminosity scenario where statistical uncertainties become smaller. Since the present work is intended as a first constraint extraction, the incorporation of experimental systematics is left for future work.

In addition, dimension-six effects are included only for the selected signal 
topology in the baseline reweighting procedure, while the background samples are kept at their SM predictions. Operators modifying $Z$--fermion couplings may also affect electroweak backgrounds such as $ZZ$ or $Z$+heavy-flavor production. A full SMEFT interpretation should include these effects consistently together with irreducible interferences among electroweak amplitudes.

\section{Conclusions}

In this work, we have investigated the \ensuremath{\mu^+\mu^-b\bar b} final state in the reconstructed \ensuremath{Z}-pole region as a probe of selected SMEFT operators. The analysis focuses on flavor-resolved interactions involving second-generation leptons and third-generation quarks, including both four-fermion operators and operators that modify effective \ensuremath{Z}--fermion couplings.

Signal and relevant background processes were simulated at leading order using \texttt{MadGraph5\_aMC@NLO}, followed by parton showering and hadronization with \texttt{Pythia8}, and fast detector simulation with \texttt{Delphes}. The event selection includes muon trigger requirements, lepton isolation, \ensuremath{b}-jet identification, missing-energy suppression, and angular separation cuts. The reconstructed invariant mass of the \ensuremath{\mu^+\mu^-b\bar b} system is used to define a signal region around the \ensuremath{Z} pole and to extract constraints 
on possible SMEFT effects.

The dependence of the selected event yield on individual Wilson coefficients was obtained using the SMEFT reweighting procedure. For each operator, the yield was parametrized as a quadratic function of the Wilson coefficient, including both the SM--dimension-six interference and the quadratic dimension-six contribution. Using an Asimov likelihood-ratio method, we derived expected 95\% C.L. constraints on the selected operators at integrated luminosities of \ensuremath{138~\mathrm{fb}^{-1}} and \ensuremath{3000~\mathrm{fb}^{-1}}.

The results show that the \ensuremath{\mu^+\mu^-b\bar b} final state can provide useful channel-specific constraints on flavor-resolved SMEFT interactions, especially on operators involving muons and bottom quarks. These constraints are complementary to those obtained from inclusive electroweak measurements, purely leptonic \ensuremath{Z} decays, and global SMEFT analyses. The projected HL-LHC luminosity further improves the reach, illustrating the potential value of including mixed leptonic--hadronic \ensuremath{Z}-pole final states in future SMEFT studies.

The present analysis provides a baseline phenomenological study of this channel. Several refinements can be incorporated in future work, including higher-order QCD and electroweak corrections, scale and PDF uncertainties, experimental systematic uncertainties, correlations among multiple Wilson coefficients, SMEFT-induced modifications of the total \ensuremath{Z} width, and a more complete treatment of electroweak interference effects in the irreducible \ensuremath{\mu^+\mu^-b\bar b} final state. These extensions would allow the constraints from this channel to be embedded more directly into global SMEFT fits and precision electroweak interpretations.

\begin{acknowledgments}
This work is supported in part by the National Natural Science Foundation of China under Grants No. 12325504.
\end{acknowledgments}

\bibliographystyle{unsrt} 
\bibliography{apssamp}% Produces the bibliography via BibTeX.

\end{document}